\documentclass{elsart}
\usepackage{amsmath}
\usepackage{amssymb}
\usepackage{graphics}
\begin{document}

\begin{frontmatter}
\title{Spin asymmetry for the $^{16}$O($\vec{\gamma},\pi^- p$)  
reaction in the $\Delta$(1232) region within an effective Lagrangian approach}

\author[CTP]{C. Fern\'andez-Ram\'{\i}rez},
\ead{cefera@mit.edu}
\author[UCM]{M.C.  Mart\'{\i}nez},
\ead{cristina@nuc2.fis.ucm.es}
\author[GSI]{Javier R. Vignote},
\author[UCM]{J.M. Ud\'{\i}as}

\address[CTP]{Center for Theoretical Physics,
Laboratory for Nuclear Science and Department of Physics, Massachusetts Institute of Technology,
77 Massachusetts Ave., Cambridge, MA 02139, USA}
\address[UCM]{Grupo de F\'{\i}sica Nuclear,
Departamento de F\'{\i}sica At\'omica, Molecular y Nuclear,
Facultad de Ciencias F\'{\i}sicas, Universidad Complutense de Madrid,
Avda. Complutense s/n, E-28040 Madrid, Spain}
\address[GSI]{Gesellschaft f\"ur Schwerionenforschung mbH, D-64291 Darmstadt, Germany}

\begin{abstract}
The spin asymmetry of the photon in the exclusive A($\vec{\gamma}$,$\pi N$)A-1
reaction is computed employing a recently developed fully relativistic model 
based on elementary pion production amplitudes 
that include a consistent treatment of the spin-3/2 nucleon resonances. We 
compare the results of this model to the only available data on 
Oxygen [Phys. Rev. C 61 (2000) 054609] and find that, contrary to other models,
the predicted spin asymmetry compares well to the 
available experimental data in the $\Delta$(1232) region. 
Our results indicate that no major medium modifications in the 
$\Delta$(1232) properties are needed in order to describe the measured 
spin asymmetries.
\end{abstract}

\end{frontmatter}


\section{Introduction}

The excitation of nucleon resonances embedded in nuclei has become an important research topic
during last decades. Among all of them, the excitation of the $\Delta$(1232) ($\Delta$ in what follows) is of 
particular relevance in nuclear reactions at intermediate energies. The possible modifications of the 
properties of the $\Delta$ during its propagation and decay  within the surrounding  medium remains an open question.

Although pion-induced reactions such as $(\pi,\pi')$ or $(\pi,\pi'N)$ were primarily invoked to shed 
light on this issue~\cite{pionscat}, the cleanest way to study both the nucleon and its excitations is through 
electromagnetic probes, i.e., photons and electrons, whose interaction with matter is better known.
Additionally, real or virtual photon-induced reactions are intrinsically much weaker than pion-induced ones, 
and can therefore sample the entire nuclear volume. 
In the last years pion photoproduction from the nucleon has focused the attention of diverse
experimental~\cite{experiments,Said} and theoretical groups~\cite{theory,Fer06a,Fer06b,update} 
worldwide,
what has allowed a good description of the $\Delta$ region.
All this research has pushed our knowledge on the low-lying resonance region to a point
where a reliable extension of such studies from free to bound nucleons is feasible.
The relevant observables for pion photoproduction off nuclei at the appropriate energies should, in principle, 
contain information on the medium modifications (if any) of the $\Delta$. Two requirements are needed before final 
conclusions can be drawn: high precision data and reliable theoretical models with
 proper $\Delta$-excitation content. The comparison of theory and data should provide the clue.
If the reaction model reproduces the data when using the $\Delta$ properties deduced from pion production from free nucleons, 
then medium modifications of the $\Delta$ are either small or they have no influence on pion production observables.
On the contrary, if the data cannot be explained by means of a reaction model with the
same $\Delta$ parameters employed in the pion production from free nucleons,
it may constitute a signature of medium modifications of the properties of the $\Delta$.
Of course, conclusions depend strongly on the reliability of the 
ingredients of the model, for instance the
nuclear description and the elementary pion production operator.

Among the photonuclear reactions that investigate the behavior of the $\Delta$ in the nuclear medium, one of the most 
interesting is the exclusive A($\gamma,\pi N$)A-1 reaction, where only one final state 
is involved. During the past 20 years, this reaction has been the focus of experiments at many facilities, 
such as TOMSK~\cite{TOMSK}, MIT-Bates~\cite{Pham92}, MAMI~\cite{MacKenzie95}, LEGS~\cite{Hicks00}, 
and NIKHEF~\cite{vanUden98}.  A non-sparse data set has been collected for double and triple differential 
cross sections, including not only $(\gamma,\pi^- p)$ but also $(\gamma,\pi^+ n)$ data, that should 
provide stringent constraints on theoretical models.
These data have been compared to calculations ranging from factorized models
-- inspired in the Blomqvist and Laget pion photoproduction model off nuclei \cite{Laget} --
to more sophisticated
distorted wave impulse approximation models  \cite{Li93,Johansson94,Lee99}.
From a careful review of the literature, one realizes that although most models succeed in reproducing partial 
sets of cross section data, there is no model capable of describing adequately the whole set of pion photoproduction 
data on nuclei. As pointed out in~\cite{Johansson94}, a major concern arises when one realizes that 
the theoretical models~\cite{Li93,Johansson94} differ strongly even at the plane-wave limit. Before 
inferring signatures of medium modifications of the $\Delta$ from these reactions, it is mandatory to 
count first on reliable calculations at least at the plane wave level.
There is a need to review the theoretical models for pion photoproduction off nuclei before progress in the 
knowledge of the in-medium $\Delta$ properties can be achieved.

Particularly interesting are the spin asymmetry data obtained at LEGS for the $^{16}$O($\vec{\gamma},\pi^- p$) reaction. 
The asymmetry is free from normalization problems, is predicted to be large, and is relatively insensitive to ambiguities in the 
theory, such as description of nonlocal effects or width of the $\Delta$ resonance~\cite{Li93}. 
In addition, the spin asymmetry is almost independent
of the pion and nucleon distortions~\cite{Li93}.
Thus, this observable becomes an excellent test for the accuracy of the underlaying
elementary pion photoproduction operator
and provides a stringent test for theoretical models.
Indeed, it was pointed out in~\cite{Li93} that 
if an experiment finds deviations of the spin asymmetry even from the simple plane-wave predictions,
this could be an indication of medium modifications of the $\Delta$ propagator.
The data collected at LEGS showed that the measured asymmetries were consistently below the theoretical predictions 
 by the Li, Wright, and Bennhold's model~\cite{Hicks00,Li93}. It was claimed that modifications to the properties 
of the $\Delta$ resonance could be necessary to achieve agreement between data and calculations~\cite{Hicks00}. 
However, this model used harmonic oscillator wave functions to describe the bound nucleon. Before definite conclusions 
are made about medium modifications of the $\Delta$, an improvement of the model ingredients, such as the struck nucleon 
wave functions and $\Delta$ Lagrangian, must be done.

In this Letter we present a model for the exclusive $A (\vec{\gamma},\pi N)$A-1 reaction, starting from the elementary 
process involving the photon, pion, nucleon and its resonances.
We perform a non-factorized computation based on a recently developed relativistic pion
photoproduction operator~\cite{Fer06a}. For free nucleons, the model developed in~\cite{Fer06a} provides a good 
description~\cite{update} of the latest fit to the world database of electromagnetic multipoles~\cite{Said}.
It is based upon an effective Lagrangian approach, fully relativistic,
and it displays gauge invariance, chiral symmetry, and crossing symmetry as well as a consistent
treatment of the spin-3/2 resonances which overcomes pathologies in former models 
\cite{Fer06a,Fer06b,Pas98}. 
The consistent treatment of the $\Delta$ should be emphasized as we intend to
look for in-medium modifications of the $\Delta$ properties.
In this Letter we apply the model only in the $\Delta$ region, however it can be applied
in further energy regions, approximately up to 1.2 GeV of photon energy.
 The extension of the model to the nucleus is introduced by means of the impulse approximation (IA), 
as described later on. 
As a first approximation one can
assume that the final state interactions (FSI) of the outgoing pion and nucleon with 
the residual nucleus can be neglected. In this case, both particles are described 
as plane waves, and one talks of the relativistic plane-wave impulse approximation 
(RPWIA)~\cite{RPWIApapers}.  
To obtain a reliable computation of the differential cross sections, the inclusion of
FSI is mandatory, but as previously stated, the spin asymmetry
can be reliably computed within RPWIA due to its low dependence on distortion effects.
In this Letter we focus on this last observable.
We show RPWIA results in the $\Delta$ region for $^{16}$O compared
to experimental data from LEGS~\cite{Hicks00}. 
We do not consider medium modifications in the nucleon resonances and we obtain
better agreement with experimental data than that the one obtained in
\cite{Hicks00} from both quantitative and qualitative points of view. 
These results indicate that major
modifications of $\Delta$ properties in the nuclear medium are not necessary
for the description of the spin asymmetry in the $^{16}$O($\vec{\gamma},\pi^- p$) process.

\section{The model}

\subsection{Relativistic Impulse Approximation}\label{sec2.1}

\begin{figure}
\rotatebox{0}{\scalebox{0.3}[0.3]{\includegraphics{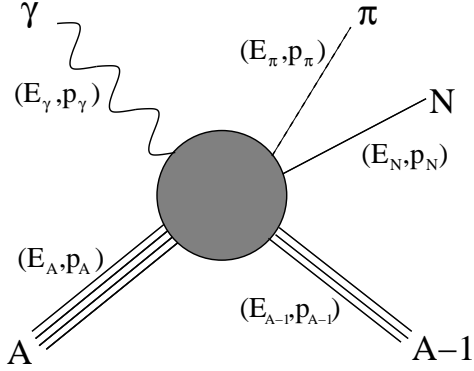}}}
\caption{Kinematics for the pion photoproduction process from nuclei.}
\label{fig:kinematics}
\end{figure}

In the exclusive A($\vec{\gamma},\pi$ N)A-1 reaction, a photon penetrates an A-body nucleus and, as a consequence 
of the interaction, a nucleon and a pion are emitted and detected, leaving behind an (A-1)-body daughter nucleus, generally 
in an excited state. The process is depicted in Fig.~\ref{fig:kinematics}, where the kinematical variables 
associated with the incoming photon and target, as well as those of the outgoing pion, nucleon, and residual nucleus 
are specified. Conservation of energy and momentum imposes that
\begin{equation}
E_{\gamma}+E_A=E_{\pi}+E_N+E_{A-1},
\end{equation}
\begin{equation}
\mathbf{p}_{\gamma}+\mathbf{p}_A=\mathbf{p}_{\pi}+\mathbf{p}_N+\mathbf{p}_{A-1}.
\end{equation}
Our calculations are performed in the laboratory frame, where the target nucleus is at rest ($E_A=M_A$, $\mathbf{p}_A=0$). 
The $z$ axis is chosen along the direction of the photon beam, and the pion is ejected in the $x-z$ plane, with azimuthal 
angle $\phi_{\pi}=0$. Although the momenta of the ejected nucleon and residual nucleus are in general not constrained 
to the $x-z$ plane, 
this coplanar kinematics, in which all the momenta in the final state belong to the same plane 
-- usually known as production plane --is experimentally the most common setup and is the
one we consider.
As can be inferred from these equations, the recoiling nucleus allows for more flexibility in the kinematics of the reaction
compared to the case of pion photoproduction from free nucleons. In fact, the three-body final state allows for the exploration 
of a wide range of momentum transfers to the residual nucleus. 

Following the conventions in~\cite{conventions},
the fivefold differential cross section for the A$(\gamma,\pi N)$A-1 reaction reads
\begin{equation}
\frac{d\sigma}{d\Omega_\pi d\Omega_N d T_N} \Big{|}_\text{lab}=
\frac{\alpha}{\left( 2\pi \right)^4}
\frac{E_N p_N p_{\pi}}{2E_\gamma}
f_{rec}^{-1} \overline{\left|  \mathcal{M}_{fi}   \right|^2},
\label{eq:difxsec}
\end{equation}
where
\begin{equation}
f_{rec} = \left| 1 - \frac{E_\pi}{E_{A-1}} \, \frac { \mathbf{p}_{A-1} \cdot
\mathbf{p}_{\pi}}{\left|\mathbf{p}_{\pi} \right|^2} \right|.
\end{equation}

The nuclear transition matrix elements for the A($\vec{\gamma},\pi$ N)A-1 reaction can be generally written as
\begin{equation}
\mathcal{M}_{fi}=\langle P_{\pi}^{\mu}, P_N^{\mu}, P_{A-1}^{\mu} | \hat{\mathcal{O}} | P_{\gamma}^{\mu}, P_A^{\mu} \rangle,
\end{equation}
where we have represented each wave function by its corresponding four-momentum. 
It is clear that for the outgoing nucleon, target and residual nucleus one must know also the spin to specify the state. 
The operator $\hat{\mathcal{O}}$ is in general an A-body operator describing the pion photoproduction process on the nucleus. 

\begin{figure}
\rotatebox{0}{\scalebox{0.3}[0.3]{\includegraphics{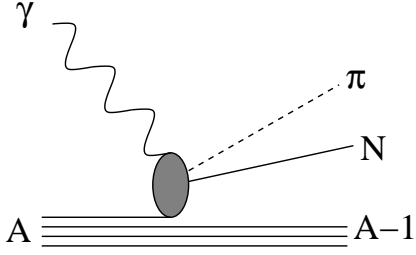}}}
\caption{Schematic representation of the impulse approximation for A($\gamma,\pi N$)A-1.}
\label{fig:ia}
\end{figure}

Our model relies on the well-known IA, as usual in processes in which the kinematics favors the interaction of the probe with a 
single nucleonic constituent of the target. To be consistent, we restrict ourselves to the quasifree region, where the momentum 
transferred to the recoiling nucleus is relatively low (below 300 MeV/c). Within the IA, the general process 
shown in Fig.~\ref{fig:kinematics} is described as illustrated in Fig.~\ref{fig:ia}, where the incoming photon interacts 
with a single bound nucleon in the nucleus. The remaining nucleons act as mere spectators in the scattering process, except for 
the FSI with both the pion and the nucleon while leaving the nucleus. It can be proven that within the IA, where the nuclear 
operator $\hat{\mathcal{O}}$ is substituted by a sum of one-body operators, the calculation of $\mathcal{M}_{fi}$ is 
simplified, and the basic ingredients that enter now in the calculation are the bound nucleon wave function, 
the elementary pion photoproduction operator, and the 
outgoing pion and nucleon wave functions.

 In our model, all of the ingredients are fully relativistic. For the elementary pion photoproduction operator, we use the 
free production operator as it is described in next section. In this work we only consider pion production from $^{16}$O, 
where a mean field description of the nuclear states is appropriate. The bound-nucleon wave function is a 
solution of the Dirac equation with well-defined angular momentum obtained in the Hartree approximation 
to the $\sigma$-$\omega$ model including non-linear $\sigma$ terms~\cite{Serot}.
We employ the NLSH wave functions by Sharma et al. \cite{Sharma} which reproduce accurately binding energies, single-particle energies, and charge radius for $^{16}$O.
As we explained in the Introduction,
we restrict ourselves to an RPWIA computation of the $^{16}$O($\vec{\gamma},\pi^- p$) spin asymmetry.
A very common theoretical framework to pion photoproduction in the nuclear medium is the use of the factorization approximation, 
that can be applied either at the amplitude or cross section levels. In a factorized calculation, the matrix elements or 
the cross sections are separated into a part containing the elementary pion production process and a part with 
the typical medium mechanisms in the process under study, such as FSI. Within a fully relativistic formalism, factorization 
is not reached even in the RPWIA, due to the presence of negative-energy contributions in the bound-nucleon 
wave function~\cite{RPWIApapers}. Thus, our calculations are fully unfactorized even in this first stage where FSI are neglected.

\subsection{Elementary pion photoproduction reaction}\label{sec2.2}
\begin{figure}
\rotatebox{0}{\scalebox{0.65}[0.65]{\includegraphics{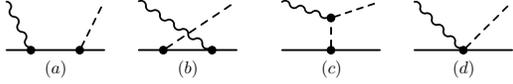}}}
\caption{Feynman diagrams for the Born terms of the
pion photoproduction from the nucleon process: ($a$) s-channel, 
($b$) u-channel, ($c$) t-channel, 
and ($d$) Kroll-Rudermann.}\label{fig:feyndiag1}
\end{figure}

\begin{figure}
\rotatebox{0}{\scalebox{0.65}[0.65]{\includegraphics{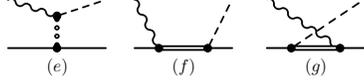}}}
\caption{Feynman diagrams for vector meson exchange ($e$) and 
resonance excitations: ($f$) s-channel and ($g$) 
u-channel  of the
pion photoproduction from the nucleon process.}
\label{fig:feyndiag2}
\end{figure}

The elementary reaction model we employ is the one developed in
\cite{Fer06a,Fer06b}
which has been applied successfully from threshold up to 1.2 GeV of
photon energy in the laboratory reference system~\cite{update}
and has been recently applied also to eta photoproduction
\cite{Fer07a}. In this section we provide a brief outlook of the
model. For further details we refer the reader to
Refs.~\cite{Fer06a,Fer06b,update}.

The model is based upon an effective Lagrangian approach (ELA) which,
from a theoretical point of view, is a very appealing, reliable, and
formally well-established approach in the energy region of the mass of
the nucleon. The model includes Born terms (diagrams (a)-(d) in Fig.
\ref{fig:feyndiag1}), vector-meson exchanges ($\rho$ and $\omega$,
diagram (e) in Fig.~\ref{fig:feyndiag2}), and all the four star
resonances in Particle Data Group (PDG)~\cite{PDG2006} up to 1.8 GeV
and up to spin-3/2: $\Delta$, N(1440), N(1520), N(1535), $\Delta$(1620),
N(1650), $\Delta$(1700), and N(1720) (diagrams (f) and (g) in Fig. 
\ref{fig:feyndiag2}).  Born terms are calculated using the Lagrangian:
\begin{equation}
\begin{split}
  {\mathcal L}_{Born}=&-ieF_1^V \hat{A}^\alpha \epsilon_{jk3}\pi_j
  \left( \partial_\alpha \pi_k \right) \\
  &-e\hat{A}^\alpha F_1^V \bar{N} \gamma_\alpha \frac{1}{2}
  \left( F_1^{S/V}+\tau_3 \right) N \\
  &-ieF_1^V\frac{f_{\pi N}}{m_\pi} \hat{A}^\alpha \bar{N}\gamma_\alpha
  \gamma_5 \frac{1}{2}[\tau_j,\tau_3] \pi_j N \\
  &-\frac{ie}{4M} F_2^V \bar{N} \frac{1}{2} \left( F_2^{S/V}+ \tau_3
  \right) \gamma_{\alpha \beta} N F^{\alpha \beta} \\
  &+\frac{f_{\pi N}}{m_\pi}\bar{N}\gamma_\alpha \gamma_5 \tau_j N
  \left( \partial^\alpha \pi_j \right) ,
\end{split} \label{eq:PVlagrangian}
\end{equation}
where $e$ is the absolute value of the electron charge, 
$m_\pi$ the mass of the pion, $M$ the mass of the nucleon,
$f_{\pi N}$
the pion-nucleon coupling constant, $F_j^V = F_j^p - F_j^n$ and $F_j^S
= F_j^p + F_j^n$ are the isovector and isoscalar nucleon form factors,
$F^{\mu \nu}=\partial^\mu \hat{A}^\nu-\partial^\nu \hat{A}^\mu$ is the
electromagnetic field ($\hat{A}^\mu$ stands for the photon field), $N$
the nucleon field, and $\pi_j$ the pion field.  The coupling to the
pion has been chosen pseudovector in order to ensure the correct
parity and low-energy behavior.

The main contribution of mesons to pion photoproduction is given by
$\rho$ (isospin-1 spin-1) and $\omega$ (isospin-0 spin-1) exchange.
The phenomenological Lagrangians which describe vector mesons are:
\begin{eqnarray}
{\mathcal L}_\omega &=& -F_{\omega N N}\bar{N}\left[\gamma_\alpha 
-i\frac{K_\omega}{2M}\gamma_{\alpha \beta}\partial^\beta \right]
\omega^\alpha N \nonumber \\
&&+\frac{e G_{\omega \pi \gamma}}{2m_\pi} 
\epsilon_{\mu \nu \alpha \beta} F^{\alpha \beta}
\left( \partial^\mu \pi_j \right) \delta_{j3}
\omega^\nu , \\
{\mathcal L}_\rho &=& -F_{\rho N N}\bar{N}\left[\gamma_\alpha 
- i\frac{K_\rho}{2M}\gamma_{\alpha \beta}\partial^\beta \right]
\tau_j \rho^\alpha_j N \nonumber \\
&&+\frac{e G_{\rho \pi \gamma}}{2m_\pi} 
\epsilon_{\mu \nu \alpha \beta} F^{\alpha \beta}
\left(\partial^\mu \pi_j \right)\rho^\nu_j.
\end{eqnarray}

The model displays chiral symmetry, gauge invariance, and crossing
symmetry as well as a consistent treatment of the spin-3/2 interaction
which overcomes pathologies present in former analyses~\cite{Pas98}.
Under this approach for spin-3/2 interactions the (spin-3/2
resonance)-nucleon-pion and the (spin 3/2 resonance)-nucleon-photon
vertices have to fulfill the condition $q_\alpha {\mathcal O}^{\alpha
  ...}=0$ where $q$ is the four-momentum of the spin-3/2 particle,
$\alpha$ the vertex index which couples to the spin-3/2 field, and the
dots stand for other possible indices. In particular, for the
$\Delta$, the simplest interacting $\pi$-$N$-$\Delta$
Lagrangian is~\cite{Pas98}
\begin{equation}
{\mathcal L}_{\pi N \Delta}=-\frac{h}{f_\pi M_\Delta} \bar{N} \epsilon_{\mu 
\nu \lambda \beta} \gamma^\beta \gamma^5 \left( \partial^\mu 
\Delta^\nu_j \right) \left( \partial^\lambda \pi_j  \right) + 
\text{H.c.} ,\label{GIcoupling}
\end{equation}
where $\text{H.c.}$ stands for Hermitian conjugate, $h$ is the strong
coupling constant, $f_\pi=92.3$ MeV is the leptonic decay constant of
the pion, $M_\Delta$ the mass of the $\Delta$, and
$\Delta^\nu_j$ the $\Delta$ field.  The
$\gamma$-$N$-$\Delta$ interaction can be written~\cite{Pas03}:
\begin{equation}
{\mathcal L}_{\gamma N \Delta}
=\frac{3e}{2M M_+} \bar{N} \left[ \frac{ig_1}{2}
\tilde{F}_{\mu \nu} +g_2 \gamma^5 F_{\mu \nu} \right] 
\left( \partial^\mu \Delta^\nu_3\right)+ \text{H.c.},
\end{equation}
where $g_1$ and $g_2$ are the electromagnetic coupling constants,
$M_+= M+M_\Delta$, and $\tilde{F}_{\mu \nu}=\epsilon_{\mu \nu \alpha
  \beta} F^{\alpha \beta}$.

The dressing of the resonances is considered by means of a
phenomenological width which contributes to both s and u channels and
takes into account decays into one $\pi$, one $\eta$, and two $\pi$.
The energy dependence of the width is chosen phenomenologically as
\begin{equation}
\Gamma \left(s,u \right) 
= \sum_{j=\pi , \pi \pi , \eta} 
\Gamma_j X_j \left( s , u \right) \: ,\label{eq:width}
\end{equation}
where $s$ and $u$ are the Mandelstam variables and
\begin{equation}
X_j \left( s , u \right) \equiv X_j \left( s \right) +  X_j 
\left( u \right) - X_j \left( s \right)  X_j \left( u \right) ,
\label{eq:Xj}
\end{equation}
with $X_j\left( l \right)$ given by
\begin{equation}
X_j \left( l \right) = 2 \frac{\left(\frac{ k_j    }{  k_{j0} } 
\right)^{2L+1}}{1+\left( \frac{ k_j  }{  k_{j0} }
\right)^{2L+3}} \: \Theta \left( l - \left( M + m_j \right)^2 \right) ,
\end{equation}
where $L$ is the angular momentum of the resonance, $\Theta$ is the
Heaviside step function, and
\begin{equation}
k_j=\sqrt{\left(l-M^2-m_j^2 \right)^2-4m_j^2M^2}/
\left( 2 \sqrt{l}\right) ,
\end{equation}
with $m_{\pi \pi} \equiv 2m_\pi$ and $k_{j0} = k_j$
when $l=M^{*2}$ ($M^*$ stands for the mass of the resonance).

This parameterization has been built in order to fulfill the following
conditions
\begin{enumerate}
\item [(i)]$\Gamma = \Gamma_0$ at $\sqrt{s}=M^*$,
\item [(ii)]$\Gamma \to 0$ when $k_j \to 0$,
\item [(iii)]a correct angular momentum barrier at threshold
  $k_j^{2L+1}$,
\item [(iv)]crossing symmetry.
\end{enumerate}

For the resonance-pion-nucleon vertex, the form factor
$\sqrt{X_\pi\left( s,u \right)}$ has to be used for consistency with
the width employed.

In order to regularize the high-energy behavior of the model, a
crossing symmetric and gauge invariant form factor is included for
Born and vector meson exchange terms,
\begin{equation}
\begin{split}
  \hat{F}_B(s,u,t)=& F(s)+F(u)+G(t)-F(s)F(u) \\
  &- F(s)G(t)-F(u)G(t)+F(s)F(u)G(t) ,
\end{split}
\end{equation}
where
\begin{eqnarray}
F(l)&=& \left[1+ \left( l-M^2 \right)^2/\Lambda^4 \right]^{-1}, 
\quad l=s,u \\
G(t)&=& \left[1+ \left( t-m_\pi^2 \right)^2/\Lambda^4 \right]^{-1} .
\end{eqnarray}

For vector mesons $\hat{F}_V(t) = G(t)$ is adopted with the change
$m_\pi \to m_V$.
In the pion photoproduction model from free nucleons
\cite{Fer06a,Fer06b} it was assumed that FSI factorize and can be
included through the distortion of the $\pi N$ final state wave
function (pion-nucleon rescattering).  $\pi N$-FSI was included by
adding a phase $\delta_{FSI}$ to the electromagnetic multipoles.  This
phase is set so that the total phase of the multipole matches the
total phase of the energy dependent solution of SAID~\cite{Said}.  In
this way it was possible to isolate the contribution of the bare diagrams
to the physical observables.  The parameters of the resonances were
extracted from data fitting the electromagnetic multipoles from the
energy independent solution of SAID~\cite{Said} applying a modern
optimization technique based upon genetic algorithms combined with
gradient based routines~\cite{Fer06b,genetico} which provides reliable
values for the parameters of the nucleon resonances.  Once the bare
properties of the nucleon resonances have been extracted from data,
their contribution to more complex problems, such as pion photoproduction from nuclei, 
can be calculated.

\section{Results}

\begin{figure*}
\rotatebox{0}{\scalebox{0.75}[0.75]{\includegraphics{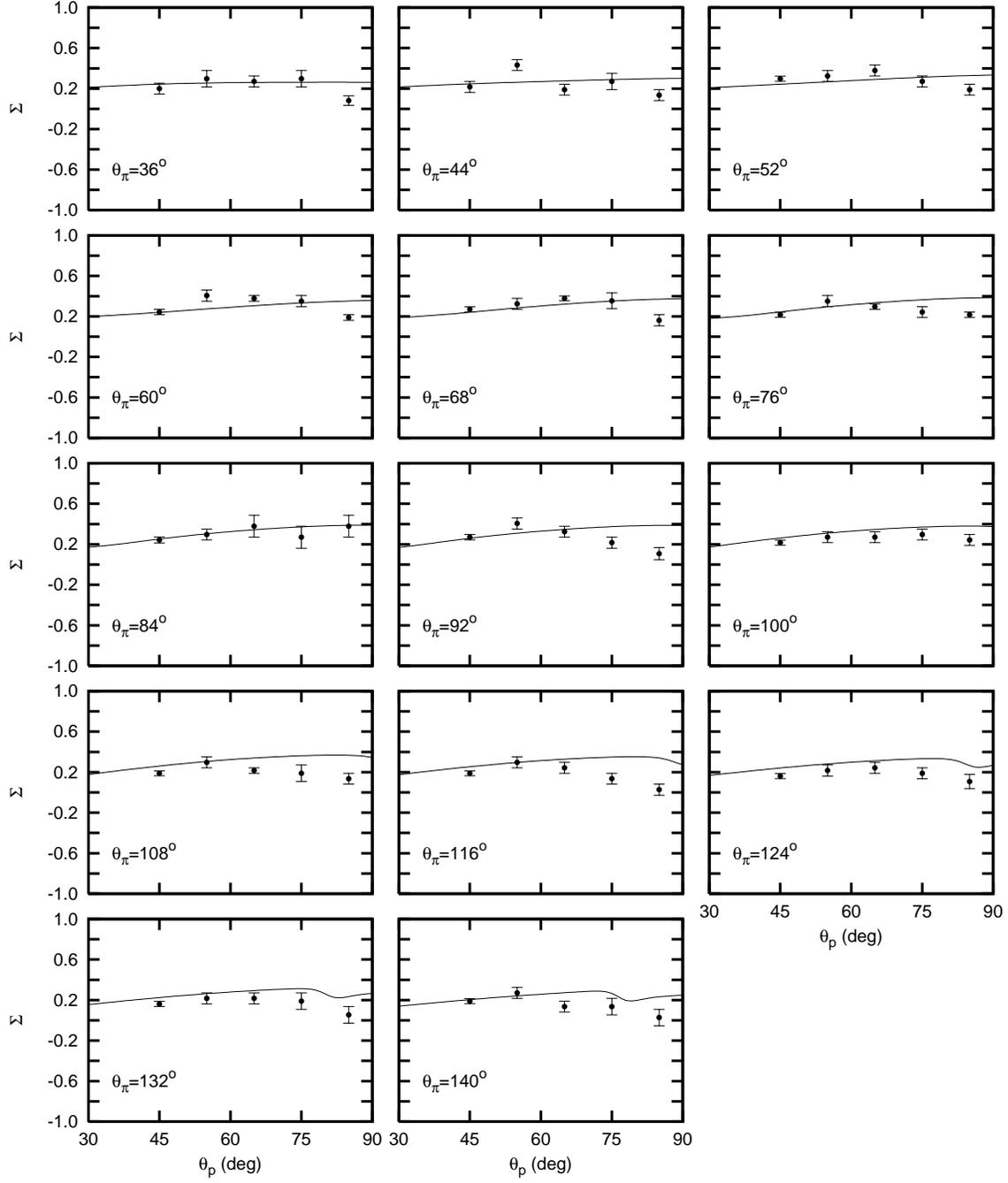}}} 
\caption{Spin asymmetry integrated over the range $T_N =$ 50 MeV to 100 MeV.
Experimental data from~\cite{Hicks00} compared to the theoretical prediction.
Combined results for the $s_{1/2}$, $p_{1/2}$, and $p_{3/2}$ shells.}
\label{fig:O16-asym}
\end{figure*}

In this section we compare the predictions of our model to the available spin asymmetry data. This asymmetry, here noted as $\Sigma$, is given by:
\begin{equation}
\Sigma = \frac{\sigma(\theta_\pi,\theta_p)_\perp - \sigma(\theta_\pi,\theta_p)_\parallel}
{\sigma(\theta_\pi,\theta_p)_\perp + \sigma(\theta_\pi,\theta_p)_\parallel},
\end{equation}
where the subindices $\perp$ and $\parallel$ stand for the perpendicular
and parallel  photon polarizations respectively and $\sigma(\theta_\pi,\theta_p)_{\perp,\parallel}$
is obtained by integrating over the nucleon kinetic energy:
\begin{equation}
\sigma(\theta_\pi,\theta_p)_{\perp,\parallel} \equiv \frac{d\sigma_{\perp,\parallel}}{d\Omega_\pi d\Omega_p} = \int\frac{d\sigma_{\perp,\parallel}}{d\Omega_\pi d\Omega_N d T_N} dT_N . \label{eq:sigma}
\end{equation}
Precise measurements of $\Sigma$ for the $^{16}$O($\vec{\gamma},\pi^- p$)
reaction at incident photon energies between 290 and 325 MeV were carried 
out at LEGS and reported in 
Ref.~\cite{Hicks00}. Data were provided at proton angles of 55$^o$ and 75$^o$ and pion angles from 36$^o$ to 140$^o$ in 8$^o$ 
steps for the sake of facilitating the comparison with theoretical calculations by preventing the need of kinematical averagings. 
In this work, we compare our theoretical predictions to those data. Our calculations include contributions 
from both $s_{1/2}$, $p_{1/2}$, and $p_{3/2}$ shells in Oxygen, consistently with the experimental setup.
The integration over the nucleon kinetic energy in Eq. (\ref{eq:sigma}) is done numerically within the same range as for the 
above mentioned $\Sigma$ data, i.e., $T_N \in [50,100]$ MeV. Our results for different pion angles $\theta_{\pi}$ as a 
function of the proton angle $\theta_p$ are shown in Fig.~\ref{fig:O16-asym}, where also the data have been plotted. 
The presentation of this figure follows the one in Ref.~\cite{Hicks00}, thus a straightforward comparison with what 
is shown in that work can be made.

As can be seen in Fig.~\ref{fig:O16-asym}, our theoretical predictions provide in general a rather good description of the data 
both from the qualitative and quantitative points of view, although the comparison worsens slightly with increasing pion and 
nucleon angles. The agreement between theory and experiment is a clear improvement with respect to what was observed 
in Ref.~\cite{Hicks00}, where it was found that the theoretical calculations based on the model in~\cite{Li93} lied 
systematically above the measured spin asymmetries (mainly for $\theta_p >$ 60$^o$). The agreement of our calculations 
with data is presumably attributed to a better description of the underlaying photon-nucleus interaction, including 
the elementary pion photoproduction operator and struck nucleon wave functions. We thus find no indication of 
$\Delta$ medium modifications in the spin asymmetry as was suggested in Ref.~\cite{Hicks00}. Of course 
the absence of in-medium effects in $\Sigma$ cannot be claimed as an absence of in-medium effects in the $\Delta$. 
One has to be cautious and has to notice that what can be claimed is that the spin asymmetry does not seem to be sensitive 
to these effects, if any.

\begin{figure}
\rotatebox{-90}{\scalebox{0.25}[0.33]{\includegraphics{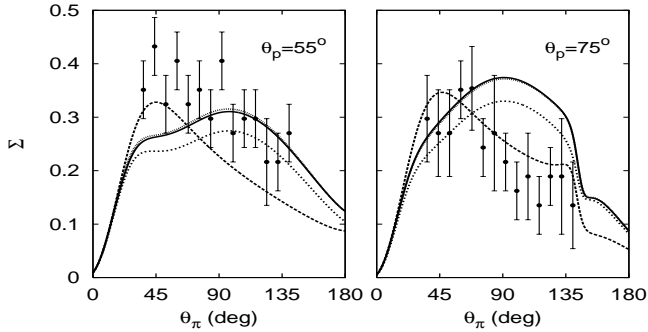}}} 
\caption{Spin asymmetry integrated over the range $T_N =$ 50 MeV to 100 MeV.
Combined results for the $s_{1/2}$, $p_{1/2}$, and $p_{3/2}$ shells. 
Curve conventions: 
Solid: Full computation; 
Dashed: Born terms and vector mesons contributions; 
Short dashed: Born terms, vector mesons, and $\Delta$ contributions;
Dotted: Born terms, vector mesons, $\Delta$, and N(1440) contributions.
Experimental data have been taken from~\cite{Hicks00}.
Dotted and solid lines almost completely overlap.}
\label{fig:O16-asymPi}
\end{figure}

\begin{figure}
\rotatebox{-90}{\scalebox{0.3}[0.32]{\includegraphics{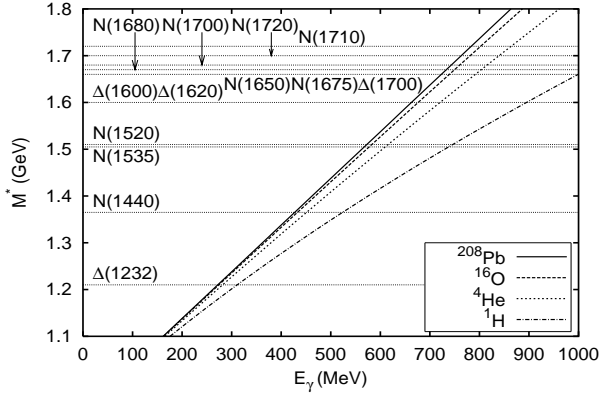}}}
\caption{Available energy for 
resonance excitations with 
different targets depending on the incident photon energy in the 
laboratory frame. Pole masses of the nucleon resonances are marked in 
the figure as horizontal lines.} \label{fig:massx}
\end{figure}

In Fig.~\ref{fig:O16-asymPi} we display the spin asymmetry computed with different contributions from the elementary 
photoproduction model. The dashed curve provides the result just accounting for Born terms and vector mesons. 
The short-dashed curve provides the calculations with Born terms, vector mesons, and $\Delta$. The dotted curve 
accounts for Born terms, vector mesons, $\Delta$, 
and N(1440) (Roper) contributions and the solid for the full computation including all the resonances.
These two last results practically overlap, what means that the contribution of higher resonances is negligible 
for the studied observables, as expected.
In the right panel it is found that Born terms and vector mesons by themselves
provide an excellent agreement with the experimental data, agreement that is spoiled
after including the $\Delta$. However, in the left panel we see that the  $\Delta$ improves agreement for 
larger pion angles. The Roper resonance shows its influence in the process although
we are in the $\Delta$ energy region. This is small but not negligible. 
It is important to notice the effect in the threshold energy for the production of the resonances
due to the fact that the knocked nucleon is bound inside the nucleus.
Indeed, when we study pion photoproduction from the nuclei, the threshold energy to produce a certain 
resonance is lowered compared to its threshold value on free nucleons. This is due to 
the fact that the whole residual system participates in the recoil so that less energy is transferred to the heavier system
and, thus, more is available to produce the resonance (see Fig.~\ref{fig:massx}).
This means that it is likely that a resonance may affect observables for lower energies than
in the free case.

We also point out the qualitative behavior of the model for high angles of both ejected pion and nucleon. 
In Fig. \ref{fig:O16-asymPi}, the asymmetry decreases slightly with increasing proton angle. The results of 
\cite{Hicks00} besides overestimating the asymmetry data, didn't follow this trend of the 
data. Our model reproduces this trend of the data, at least qualitatively. This behavior of the asymmetry
is found even for Born terms (see Fig.~\ref{fig:O16-asymPi}) and it is a kinematical effect.
When the asymmetry is compared to the mean value of the kinetic energy of the outgoing nucleon 
$\langle T_N \rangle$, it can be seen that it is  the variation of $\langle T_N \rangle$ what 
is seen in this behavior of the asymmetry.

\section{Summary and final remarks}

It has been suggested that
the spin asymmetry in A($\vec{\gamma}$,$\pi N$)A-1 reaction may serve to signal
in-medium $\Delta$ modifications.
In this paper we have presented results of a new model for pion photoproduction
on nuclei to the description of this observable in the $^{16}$O($\vec{\gamma}$,$\pi^- p$)
reaction measured at LEGS~\cite{Hicks00}. The model is an extension to nuclei of the model of
\cite{Fer06a,Fer06b,update} for free nucleons. A salient feature of the model is the improved
treatment of the spin-3/2 resonances. One must keep in mind that a consistent description of the
$\Delta$-resonance is compulsory previous to any comparison with data.
Our results within the plane-wave limit are
in fair agreement with the experimental data on the spin asymmetry. This indicates that FSI
are not significant in the description of this asymmetry, in agreement with the findings in
\cite{Li93}.
Within our model, the description of the spin asymmetry is obtained
with the same $\Delta$ parameters used to describe pion photoproduction data on free nucleons.
This result indicates that major in-medium $\Delta$ effects are not needed to reproduce asymmetry data.

\begin{ack}
C.F.-R. and M.C.M. thank Dr. L. \'Alvarez-Ruso for useful discussion.
C.F.-R. thanks Dr. T.W. Donnelly for useful comments.
The authors acknowledge partial support
from Ministerio de Educaci\'on y Ciencia (Spain) under contracts FPA2006-07393
and FPA2007-62216 and by UCM and Comunidad de Madrid
under project number 910059 (Grupo de F\'{\i}sica Nuclear).
M.C.M. is supported by the Spanish program ``Juan de la Cierva''.
C.F.-R. and J.R.V are supported by "Programa de Becas Postdoctorales" of  
Ministerio de Educaci\'on y Ciencia (Spain). 
Part of the computations of this work were carried out at the 
``Cluster de C\'alculo de Alta Capacidad para T\'ecnicas F\'{\i}sicas''
funded by EU Commission (FEDER programme) and by UCM.
\end{ack}

\end{document}